\documentclass[draftcls,onecolumn,12pt,twoside,journal]{IEEEtranTCOM}
\pdfoutput=1
\normalsize

\usepackage{cite}
\usepackage{amsmath,amssymb,amsfonts}
\usepackage{algorithmic}
\usepackage{graphicx}
\usepackage{textcomp}
\usepackage{graphicx}                                            
\usepackage{multicol}         
\usepackage{caption}
\usepackage{amssymb,amsmath,dsfont}
\usepackage{inputenc}
\usepackage{bbm}
\usepackage{comment}
\usepackage{cite}
\usepackage{color}
\usepackage{multirow}
\usepackage{subcaption}
\usepackage{array}
\usepackage{algorithm}
\usepackage{color,soul}
\usepackage{blindtext}
\usepackage{bm}
\usepackage[sort&compress,numbers]{natbib}
\usepackage{mdframed}
\setcitestyle{square}

\makeatletter
\newcommand{\mathleft}{\@fleqntrue\@mathmargin0pt}
\newcommand{\mathcenter}{\@fleqnfalse}
\makeatother

\newtheorem{proposition}{Proposition}
\newtheorem{proof}{Proof}

\begin{document}
\title{Stochastic geometry analysis of a distance-based JT scheme in C-RAN}

\author{Charles~Wiame,~\IEEEmembership{Member,~IEEE}, Claude~Oestges,~\IEEEmembership{Fellow,~IEEE}, and~Luc~Vandendorpe,~\IEEEmembership{Fellow,~IEEE}
\thanks{C. Wiame, C. Oestges and L. Vandendorpe are with the Institute of Information and Communication Technologies, Electronics and Applied Mathematics (ICTEAM), UCLouvain, Louvain-la-Neuve, Belgium.}}

\maketitle

\begin{abstract}
This paper considers a joint transmission scheme (JT) developed for cloud radio access networks (C-RANs). This proposed scheme features cooperative sets of remote radio heads (RRH) defined in a disk around each user location. The nodes belonging to each of these sets perform a weighted maximum ratio transmission to jointly  serve the user. The powers allocated to the beamformers are computed at the network baseband unit, taking into account channel gains, as well an equity criterion between the users. In comparison with the existing literature, our model includes a saturation assumption, with all transmissions taking place over the same resource block. A RRH belonging to multiple sets can hence transmit to several users simultaneously. The distributions of the network coverage and spectral efficiency are calculated by means of stochastic geometry (SG), and compared with Monte Carlo simulations. The derived expressions take into account the power allocation, the user and RRH densities, as well as the statistical correlation resulting from the set overlaps. 
\end{abstract}


\IEEEpeerreviewmaketitle

\section{Introduction}


C-RANs are considered as a promising network architecture for future mobile generations (5G and beyond) \cite{Intro_1}.\\

\begin{figure}[h!]
    \centering
    \includegraphics[width = 0.6\textwidth]{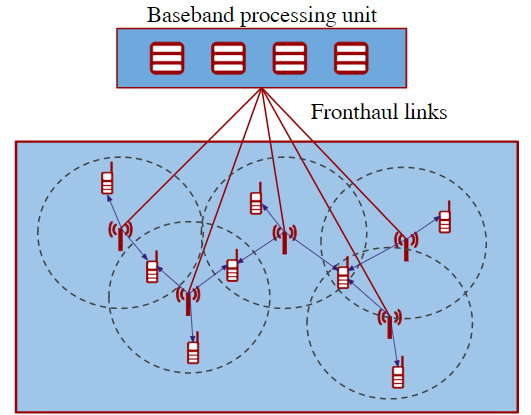}
    \caption{Simplified representation of a C-RAN.}
    \label{CRAN}
\end{figure}

As depicted in Figure \ref{CRAN}, this architecture includes RRHs connected to a baseband unit (BBU) by means of high capacity fronthaul links. A major difference with traditional cellular networks lies as a consequence the signal processing, which can be centralized and optimized at the BBU.\\

In addition, a joint optimization has to take into account the computational cost and the required overhead to acquire channel state information. One of the main challenges in the study of C-RANs is hence the development of algorithms meeting this trade-off. \\

Among the possible elements to be optimized, one can mention the resource allocation, the RRH deployment, as well as the user association policy. This paper focuses on the last aspect by analyzing a joint cooperation strategy.

\subsection{Brief overview of related works}

A large number of publications employ SG to evaluate the statistical performance of C-RANs \cite{9186615,7491366,7944682,6600939,7018059,7124520,7415954,8400544,9217353,8972478,7334848, 7425275, 7460258, 7982774, 8063963, 8422228, 9024639, 9082905, 8254187, 9050646, 7248933, 8974591, 7365885, 6860252, 9201540, 6928420, 7738565, 8100895, 8937506, 8690607, 6582745, 6824977,6851166,7414099,7450690,8289331,8422715,8526351,9247173,7448831,9145416} . In most of these works, the following association policies are studied: 
\begin{itemize}
    \item \textit{Selection transmission (ST)}: each mobile user is served by one RRH  \cite{9186615,7491366,7944682}. This RRH can be for instance the closest to the user, providing the strongest received power \cite{9186615}, or be randomly selected among a set of candidates \cite{7491366}.
    \item \textit{All RRHs participate (ARP)}: in finite networks, all the RRHs deployed in the considered area coordinate to serve the user \cite{6600939,7018059,7124520,7415954,8400544,9217353,8972478}.
    \item \textit{Partial coordination (PC)}: a subset of the existing RRHs are selected to jointly transmit to the user \cite{7334848, 7425275, 7460258, 7982774, 8063963, 8422228, 9024639, 9082905, 8254187, 9050646, 7248933, 8974591, 7365885, 6860252, 9201540, 6928420, 7738565, 8100895, 8937506, 8690607, 6582745, 6824977,6851166,7414099,7450690,8289331,8422715,8526351,9247173,7448831,9145416}.
\end{itemize}

The joint transmission (JT) scheme proposed in this work features PC: cooperation zones are defined around each user based on its location. 

\subsection{Contributions}

This paper presents a JT scheme, relying on cooperative sets of RRHs around each user. These sets are defined using a distance threshold criterion, and contain the RRHs located in disks around the users. The network is assumed to be fully loaded, i.e. all the transmissions take place over the same resources. The main contributions of this work can be summarized as follows: 

\begin{itemize}
    \item The cell free approach proposed in this paper features dynamic clustering: cooperative sets form in real time around each user position. In addition, the channel values of the RRHs located in given set are centralized at the BBU. This joint knowledge is employed to dynamically allocate the transmit powers of these RRHs.
    \item Our framework considers the most general case of $N$ coordinating RRHs, each equipped with $M$ antennas. Due to the random nature of the point processes, the number of nodes located in the cooperation zones varies from user to user. This number $N$ is hence modeled as a random variable, as additional degree of generality compared to most existing works.
    \item Unlike many existing CoMP models, all transmissions are here assumed to be performed over the same resource block. The total interference signal is hence composed of a double summation: over symbols dedicated to other users, and over the RRHs serving each of these users. The number of terms involved in the first sum is function of the user density, while the second sum is related to the RRH density. Using SG, we manage to derive performance metrics depending on these two densities. This differs from previous publications, which mostly assume TDD/FDD, therefore without dependency on the user density. To our best knowledge, \cite{8974591} is the only other work considering PC where the performance is also derived as function of both parameters.
    \item The cooperation zones can spatially overlap. Since all transmissions take place over the same slot, the user can receive both useful information and interference from a same RRH. This will be the case if this RRH is in the cooperative set of more than one user. As a result, the useful and interference power are statistically correlated due to the number and locations of the RRHs sending both of them. Unlike the approach proposed in \cite{8974591}, our SG model captures this statistical correlation.
\end{itemize}

\subsection{Organization of the paper}
The rest of this paper is organized as follows: the system model is described in section II. The analytical results derived using SG are presented in Section III. The numerical results and the conclusion are respectively detailed in sections IV and V. \\

\textit{Notations:} in the next sections, $j=\sqrt{-1}$ denotes the imaginary unit. $\operatorname{Im}\{\cdot\}$ represents the imaginary part of a complex number. Bold letters denotes vectors. $\mathbf{x}^H$ is the conjugate transpose of $\mathbf{x}$. $[\mathbf{x}_p]_{p \in \mathcal{P}}$ is the concatenation of vectors $\mathbf{x}_k$ whose index $p$ belongs to the set $\mathcal{P}$. $\mathbf{x}_k \in \mathbb{R}^2$ denotes the cartesian coordinates of a network node $k$. $\mathcal{D}(\mathbf{x},R)$ is the disk of radius $R$ centered around $\mathbf{x} \in \mathbb{R}^2$. $|\mathcal{C}|$ denotes the cardinal number of set $\mathcal{C}$.

\section{System Model}
\label{sect:system}

\subsection{Network topology}
We consider the downlink of a C-RAN deployed in $\mathbb{R}^2$. The RRHs of this network are distributed according to a homogeneous Poisson point process (HPPP) $\Psi_R$ of intensity $\lambda_R$. Given a realization of this point process, we denote $\mathbf{x}_i$, the coordinate of RRH $i \in \Psi_R$.
Each of these RRH are equipped with $M$ antennas. They are also all connected via fronthaul links to a core BBU.

Single antenna users (UE) are distributed using a second HPPP $\Psi_U$ of intensity $\lambda_U$. This point process is deployed over an area $\mathcal{A} \subset \mathbb{R}^2$ conditioned on the prior realization of $\Psi_R$. The area is defined as 
\begin{equation}
 \mathcal{A} = \mathbb{R}^2 \setminus \Big( \cup_{i \in \Psi_R} \mathcal{D}(\mathbf{x}_i,r_0) \Big)
\end{equation}
with $r_0 = 1 \; m$, an exclusion radius. As explained in the next section, this area definition enables to avoid unrealistic path loss values. Each user is equipped with a single antenna.

\begin{figure}[h!]
    \centering
    \includegraphics[width = 0.53 \textwidth]{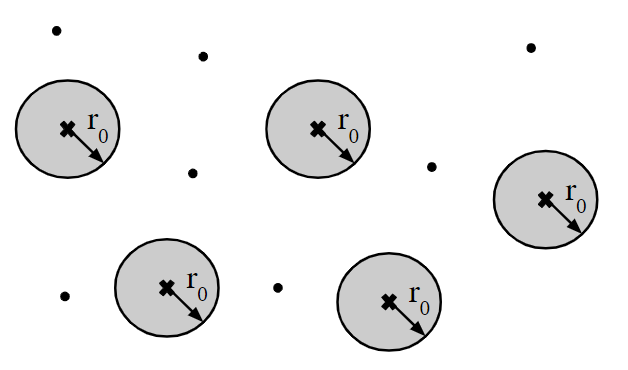}
    \caption{Illustration of the two point processes $\Psi_R$ (represented by the crosses) and $\Psi_U$ (represented by the dots).}
    \label{network}
\end{figure}

The two point processes are illustrated in Figure \ref{network}.

\subsection{Channel model}

The channel vector $\mathbf{g}_{ij} \in \mathbb{C}^{M \times 1}$ between RRH $i$ and user $j$ is given by 
\begin{equation}
    \mathbf{g}_{ij} = \mathbf{h}_{ij} r_{ij}^{-\alpha}.
\end{equation}
In this definition, the following elements have been introduced

\begin{itemize}
    \item $\mathbf{h}_{ij} \in \mathbb{C}^{M \times 1}$, a vector containing the fading coefficients $h_{ijk}$ (with $k=1,\hdots,M$). These coefficients are assumed to be independently distributed. We consider a Rayleigh fading scenario, with therefore $h_{ijk} \sim \mathcal{CN}\big(0,\frac{\sqrt{2}}{2}\big)$ and $|\mathbf{h}_{ijk}|^2 \sim \exp(1)$.
    \item $r_{ij}^{-\alpha}$ is the path loss, with $\alpha > 2$, the path loss exponent and $r_{ij}$, the Euclidean distance between $i$ and $j$. Since $\Psi_U$ is defined over area $\mathcal{A}$, the path loss does not take unrealistic values higher than 1. 
\end{itemize}

For sake of mathematical tractability with SG, shadowing effects are not modelled in this paper. 

\subsection{Association policy}
In order to limit the amount of overhead, each RRH only estimates the channels of the users located at a distance lower than a predefined radius $r_1 > r_0$. This estimation is assumed to be perfect.

Given this channel knowledge, each user $j$ is then jointly served by these RRHs located in $\mathcal{D}(\mathbf{x}_j,r_1)$ (in other words, all the RRHs $i$ having estimated the channel vector $\mathbf{g}_{ij}$). This zone $\mathcal{D}(\mathbf{x}_j,r_1)$ is denoted \textit{cooperative set} of user $j$, in the rest of this paper.

\begin{figure}[h!]
    \centering
    \includegraphics[width = 0.55 \textwidth]{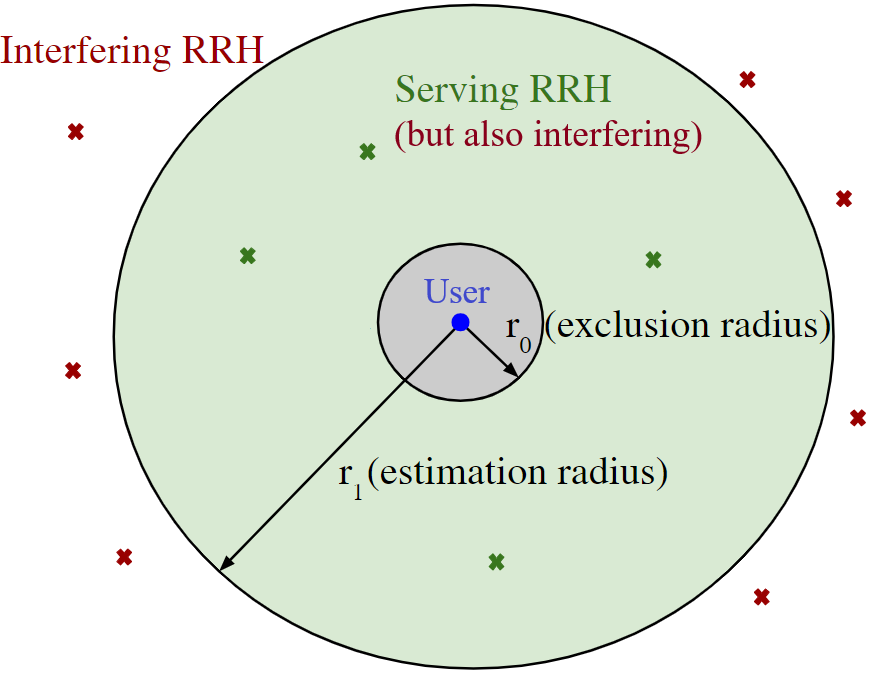}
    \caption{A given user (represented in blue) has no RRH around it at a distance lower than $r_0$ (due to the definition of $\Psi_U$). This user is served by the green RRHs located in its estimation zone. These green RRHs can simultaneously transmit interference since they can serve other users (not represented in the figure). The red RRH located beyond $r_1$ transmits interference only.}
    \label{Association}
\end{figure}

Due to the random node distribution, the different cooperative sets can overlap: a given RRH can thus be associated to several users. As previously mentioned, it is assumed that all downlink transmissions are performed over the same resource block. Under this assumption, a RRH can possibly transmit a superposition of signals intended to distinct users. Each of these users will in that case simultaneously receive useful information and interference from this node. This association policy is illustrated in Figure \ref{Association}.

\subsection{Signal model}

Without loss of generality, the network performance is evaluated for a centric user $j^*$ added at $(0,0)$. According to Slivnyak-Mecke theorem \cite{Baccelli}, introducing this centric user does not change the properties of the HPPP. 

The baseband signal received at the centric user is given by
\begin{equation}
    \text{y}_{j^*} = \sum_{i \in \mathcal{C}_{j^*}} \mathbf{w}_{ij^*}\mathbf{h}_{ij^*} r^{-\alpha/2}_{ij^*} s_{j^*} + \sum_{j \in \Psi_{U} \backslash \{j^*\}} \Bigg(\sum_{i \in \mathcal{C}_{j}} \mathbf{w}_{ij}\mathbf{h}_{ij^*} r^{-\alpha/2}_{ij^*} \Bigg) s_j + n_{j^*}
\end{equation}

where the following notations have been introduced

\begin{itemize}
    \item $\mathcal{C}_{j}$ is the set of RRHs serving user $j$;
    \item $s_j$ is the unit energy symbol intended for user $j$;
    \item $\mathbf{w}_{ij}$ is the beamforming vector associated to transmission from $i$ to $j$.
    \item $n_j$ is the additive thermal noise of constant power $N_0$.
\end{itemize}

The power received at the centric user is therefore given by 
\begin{equation}
\label{eqn_power}
\begin{split}
    P_{j^*} &  = \underbrace{\bigg| \sum_{i \in \mathcal{C}_{j^*}} \mathbf{w}_{ij^*}\mathbf{h}_{ij^*} r^{-\alpha/2}_{ij^*} \bigg|^2}_{P_{U}} + \underbrace{\sum_{j \in \Psi_{U} \backslash \{j^*\}} \bigg| \sum_{i \in \mathcal{C}_{j}} \mathbf{w}_{ij}\mathbf{h}_{ij^*} r^{-\alpha/2}_{ij^*} \bigg|^2}_{P_{I}} + \; N_0
    \end{split} 
\end{equation}
where $P_{U}$ is the useful information power and $P_{I}$, the aggregate interference.

\subsection{Beamforming strategy}

In the framework of this paper, we consider maximum ratio transmission as beamforming strategy\footnote{other traditional beamforming strategies (e.g. zero forcing or minimum mean square error combining) lead to mathematical expression that are more difficult to incorporate in a SG framework.}. We therefore have 
    \begin{equation}
        \mathbf{w}_{ij} = \dfrac{\mathbf{g}_{ij}^H}{\gamma_{ij}}
    \end{equation}
    
where $\gamma_{ij}$ is a normalization factor, which can be chosen in several manners. In the framework of this paper, this factor is defined as $\gamma_{ij} = |\mathbf{g}_{i}|$ with $\mathbf{g}_{i} = [\mathbf{g}_{ij}]_{i \in \mathcal{C}_j}$ being the concatenation of the channel vectors of all RRHs serving $j$.

This normalization enables to guarantee a constant unit power budget allocated per user. The baseband signal emitted by RRH $i$ is indeed given by $\mathbf{w}_{ij}s_j$. The total transmit power dedicated to each user $j$ can hence be express as 
\begin{equation}
    P_{tj} = \Big| \sum_{i \in \mathcal{C}_{j}} \mathbf{w}_{ij} s_j \Big|^2 = \dfrac{1}{|\mathbf{g}_{j}|^2} \sum_{i \in \mathcal{C}_{j}} |\mathbf{w}_{ij}|^2 = 1.
\end{equation}

Although the power allocated by the BBU per user is constant, it is important to note that it is nonuniformly distributed among the RRH serving it. The power allocated to RRH $i$ in order to serve user $j$ is indeed given by $|\mathbf{w}_{ij}s_j|^2 = |\mathbf{g}_{ij}|^2/|\mathbf{g}_{j}|^2$. RHH experiencing better channel conditions are therefore granted more power. Depending on chosen criterion, other normalization choices\footnote{We can mention as other normalization criteria \begin{itemize}
    \item a constraint on the norm of $\mathbf{w}_{ij}$;
    \item a budget on the total power consummed by RRH $i$;
    \item $\hdots$
\end{itemize}} might be more relevant. The study of these other options is left for future work. In our case, the above normalization ensures equity in the powers allocated to serve the users.

\section{Performance Analysis} \label{sect:analysis}

The main result of this section is the coverage probability provided in propositions \ref{GP_corr} to \ref{outside} The preliminary reasoning and assumptions required to derive this expression are detailed in propositions \ref{CT} and \ref{moment_matching}.

Considering (\ref{eqn_power}), the signal-to-interference-plus-noise-ratio (SINR) of the centric user is given by 
\begin{equation}
    \text{SINR} = \dfrac{P_{U}}{P_{I} + N_0} = \dfrac{\bigg| \sum_{i \in \mathcal{C}_{j^*}} \mathbf{w}_{ij^*}\mathbf{h}_{ij^*} r^{-\alpha/2}_{ij^*} \bigg|^2}{\sum_{j \in \Psi_{U} \backslash \{j^*\}} \bigg| \sum_{i \in \mathcal{C}_{j}} \mathbf{w}_{ij}\mathbf{h}_{ij^*} r^{-\alpha/2}_{ij^*} \bigg|^2 + N_0}.
\end{equation}

The expression of the aggregate interference $P_{I}$ is complex and has to be simplified in order to fit in a stochastic geometry framework.

\begin{proposition}
\label{CT}
The squared norm in $P_{I}$ can be developed as a sum of products. Following the approach proposed in \cite{}, the cross terms of this sum (involving non-identical indexes) are neglected. Using this approximation, the aggregate interference power received at the typical user can be rewritten as 
\medskip
\begin{equation}
        P'_{I} = \sum_{i \in \mathcal{C}_{j^*}} \; \sum_{\substack{j \in \mathcal{B}_i \\ j \neq j^*}} \;  \dfrac{\sum_{k=1}^{M} \big| h_{ijk}\big|^2 \big| h_{ij^*k} |^2 r^{-\alpha}_{ij}}{\sum_{i' \in \mathcal{C}_{j}} \sum_{k=1}^{M} \big| h_{i'jk}\big|^2 r^{-\alpha}_{i'j}} \; r^{-\alpha}_{ij^*} + \sum_{i \in \Psi_{R}\setminus \mathcal{C}_{j^*}} \; \sum_{\substack{j \in \mathcal{B}_i \\ j \neq j^*}} \;  \dfrac{\sum_{k=1}^{M} \big| h_{ijk}\big|^2 \big| h_{ij^*k} |^2 r^{-\alpha}_{ij}}{\sum_{i' \in \mathcal{C}_{j}} \sum_{k=1}^{M} \big| h_{i'jk}\big|^2 r^{-\alpha}_{i'j}} \; r^{-\alpha}_{ij^*}
\end{equation}
\medskip
where $\mathcal{B}_i$ denotes the set of users served by RRH $i$.

\proof
See appendix \ref{proof_CT}.
\end{proposition}

\medskip

The quotients of Rayleigh fading coefficients are still to complex to capture using SG. For this reason, the approximation of the next proposition is introduced. 

\medskip

\begin{proposition}
\label{moment_matching}
Let $N_j = |\mathcal{C}_{j}|$. Using moment matching, the coefficients 
\begin{equation}
\dfrac{\sum_{k=1}^{M} \big| h_{ijk}\big|^2 \big| h_{ij^*k} |^2 r^{-\alpha}_{ij}}{\sum_{i'=1}^{N_j} \sum_{k=1}^{M} \big| h_{i'jk}\big|^2 r^{-\alpha}_{i'j}}
\end{equation}

are approximated by a Gamma random variable $Z^{(i,j)}_{N_j}$ of shape and scale parameters given by 
\begin{minipage}{0.49\textwidth}
\begin{align*}
    k_{N_j} = \begin{cases}
    \frac{M^2+2M-1}{2M} & \text{if}\; N_j = 1\\
    \frac{MN_j - 1}{3MN_j(N_j-1)} & \text{if}\; N_j > 1
    \end{cases}
\end{align*}

\end{minipage}
\hfill
\hfill
\begin{minipage}{0.49\textwidth}
\begin{align*}
    s_{N_j} = \begin{cases}
    \frac{2M}{M^2+2M-1} & \text{if}\; N_j = 1\\
     \frac{3(N_j-1)}{N_j(MN_j-1)} & \text{if}\; N_j > 1
    \end{cases}
\end{align*}
\end{minipage}

\medskip
\medskip
\proof See appendix \ref{proof_moment_matching}.
\end{proposition}

\medskip

Using this proposition, the aggregate interference power can be rewritten as 
\begin{equation}
        P'_{I} = \underbrace{\sum_{i \in \mathcal{C}_{j^*}} \; \sum_{\substack{j \in \mathcal{B}_i \\ j \neq j^*}} \;  Z^{(i,j)}_{|\mathcal{C}_{j}|} r^{-\alpha}_{ij^*}}_{P_{I_1}} + \underbrace{\sum_{i \in \Psi_{R}\setminus \mathcal{C}_{j^*}} \; \sum_{\substack{j \in \mathcal{B}_i \\ j \neq j^*}} \;  Z^{(i,j)}_{|\mathcal{C}_{j}|} r^{-\alpha}_{ij^*}}_{P_{I_2}}.
\end{equation}

In this last expression, the first term $I_1$ comes from RRHs which also serve the centric user. This term depends 
\begin{itemize}
    \item on the number $|\mathcal{C}_{j^*}|$ of RRHs serving $j^*$;
    \item on the number of users $|\mathcal{B}_i|$ served by each of these RRHs $i$;
    \item on the number of RRHs $|\mathcal{C}_{j}|$ serving each of these users served by RRH $i$ (because of the normalization factor). 
\end{itemize}

\medskip
The two first items are also present in the useful information power provided to the user in (\ref{eqn_power}). Interference $I_1$ is therefore statistically correlated to the useful information power. This correlation has to be taken into account in order to accurately derive the coverage.

By contrast, interference $I_2$ comes from RRH outside the cooperative set of $j^*$. This term is therefore assumed to be independent from the useful power. 

\medskip

\begin{proposition}
\label{GP_corr}
Taking into account the correlation between the useful power and interference $I_1$, the coverage probability can be expressed as 
\begin{align}
    \label{coverage_eq}
    \mathcal{P}(\theta) &= \frac{1}{2} + \frac{1}{\pi}\sum_{n_U = 0}^{\infty}\sum_{n_R = 0}^{\infty} p(n_R)p(n_U) \int_{0}^{\infty}\operatorname{Im}\{\phi_{P_{U'}}(t|n_R,n_U) \phi_{P_{I_2}}(-\theta t) e^{-jt\theta N_0}\}t^{-1}dt.
\end{align}

with the following notations
\begin{itemize}
    \item $p(n_R)$ and $p(n_U)$, the probabilities to have $n_R$ RRHs and $n_U$ users in the cooperative set of $j^*$
    \begin{align}
        p(n_R) &= \Big[\lambda_R\pi(r_1^2-r_0^2)\Big]^{n_R}\frac{e^{-\lambda_R\pi(r_1^2-r_0^2)}}{n_R!}. \\
        p(n_U) &= \Big[\lambda_U\pi(r_1^2-r_0^2)\Big]^{n_U}\frac{e^{-\lambda_U\pi(r_1^2-r_0^2)}}{n_U!}.
        \label{pnu}
    \end{align}
    \item $P_{U'}$, a random variable defined as $P_{U} - \theta P_{I_1}$.
    \item $\phi_{P_{U'}}(t|n_R,n_U)$, the characteristic function (CF)  of $P_{U'}$, conditioned on the number of users and RRHs present in the cooperative set of $j^*$. This function is derived in Proposition \ref{corr2}. 
    \item $\phi_{P_{I_2}}(t)$, the CF of $P_{I_2}$. The expression of this function is presented in Proposition \ref{outside}.

\end{itemize}

\proof
See appendix \ref{proof_GP_corr}.
\end{proposition}

\medskip

The spectral efficiency distribution can be easily deduced from this last result. Using the classical definition $\text{SE} = \log_2(1+ \text{SINR})$, the cumulative distribution function of SE is directly given by $\mathcal{P}(\theta')$, thanks to the change of variable $\theta' = 2^{\theta} - 1$. 

\medskip

\begin{figure}[h!]
    \centering
    \includegraphics[width = 0.55\textwidth]{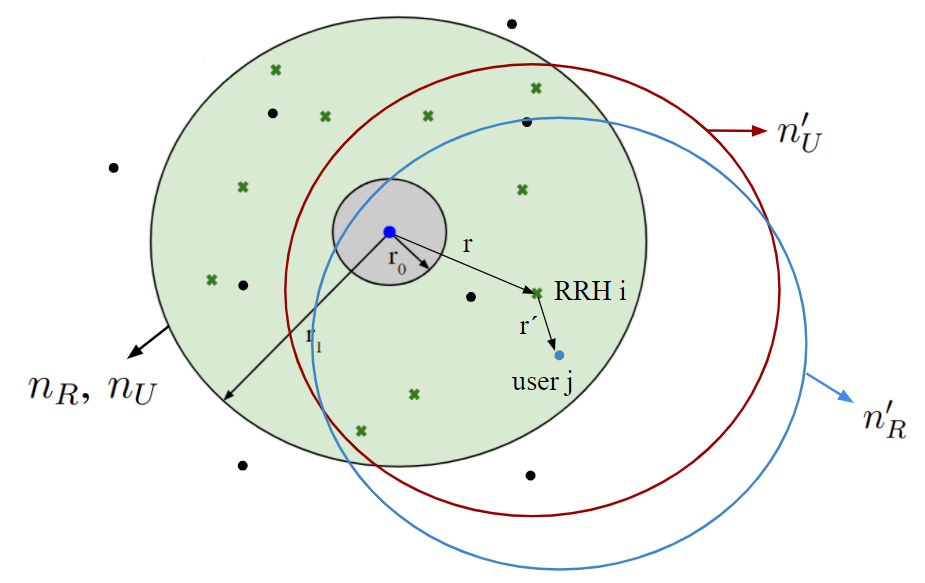}
    \caption{Spatial correlation between the cooperative sets.}
    \label{corr_circles}
\end{figure}

In order to introduce the next proposition, let us consider the elements depicted in Figure \ref{corr_circles}:
\begin{itemize}
    \item $n_R$ RRHs and $n_U$ users in the cooperative set of the centric user $j^*$;
    \item a RRH $i$ located at a distance $r\in [r_0;r_1]$ serving $j^*$;
    \item $n_U'$, the total number of users served by $i$ (contained in the red circle);
    \item $j\neq j^*$, another user served by $i$, located at a distance $r'$ from it;
    \item $n_R'$, the total number of RRH serving $j$ (contained in the blue circle).
\end{itemize}

One can observe that the numbers $n_U$ and $n_U'$ are correlated (due to the intersection of the red and green circles). The same observation can be made for $n_R$ and $n_R'$ (due to the intersection of the green and blue circles).

The objective of the following proposition is now to provide an expression of the CF of $P_{U'}$, taking into account the correlations between $P_U$ and $P_{I_1}$. 

\begin{proposition}
Conditioned on the values of $n_R$ and $n_U$, the CF of $P_{U'}$ is given by 
\begin{equation}
\label{SandT}
    \phi_{P_{U'}}(t|n_R,n_U) = \Bigg[\int_{r_0}^{r_1} \phi_{S}(t|n_R,n_U,r)\phi_{T}(-\theta t|n_R,n_U,r) \frac{2r}{r_1^2-r_0^2}\Bigg]^{n_R}
\end{equation}

where $\phi_{S}(t|n_R,n_U,r)$ and $\phi_{T}(-\theta t|n_R,n_U,r)$ are the CFs of the useful and interference power provided by one serving RRH at distance $r$ of the origin. These functions can be expressed as 
\begin{align}
    \label{phi_S}
    \phi_{S}(t|n_R,n_U,r) &= (1-jtr^{-\alpha})^{-M}\\
    \phi_{T}(t|n_R,n_U,r) &= \prod_{n_U'=0}^{\infty} p(n_U'|n_U) \bigg[\prod_{n_R' = 1}^{\infty} p(n_R'|n_R) \; \phi_{V}\Big( t\big|n_R',r \Big) \bigg]^{n_U'}.
    \label{phi_T}
\end{align}

In the last expression, $\phi_{V}\Big( t\big|n_R',r \Big)$ is the CF of the interference power coming from the signal emitted by a RRH $i$ for a user $j\neq j^*$, which is itself served by $n_R'$ users in total. This function is given by
\begin{equation}
    \label{phiV}
    \phi_{V}\Big( t\big|n_R',r \Big) = (1-jtr^{-\alpha}s_{n_R'})^{-k_{n_R'}}.
\end{equation}

The conditional probabilities $p(n_U'|n_U)$ and $p(n_R'|n_R)$ are provided in appendix
\label{corr2}

\proof
See appendix \ref{proof_corr2}.
\end{proposition}

The only function left to compute is the CF of $P_{I_2}$. Let $\Tilde{\Psi}_R = \Psi_R \setminus \mathcal{C}_{j^*}$ be the set of RRHs generating $P_{I_2}$. Each of these RRHs serves a different number of users. In order to derive a closed form result, this set is decomposed as an infinite number of HPPPs
\begin{equation}
    \Tilde{\Psi}_R = \bigcup\limits_{n=0}^{\infty} \Tilde{\Psi}_{R,n}
\end{equation}
where $\Tilde{\Psi}_{R,n}$ is the set of RRHs serving exactly $n$ users. By using the displacement theorem \cite{Baccelli}, the reduced density of each of these of these PPP is given by $\lambda_{R,n} =\lambda_{R} \; p(n)$ with $p(n)$ given by (\ref{pnu}).
\begin{proposition}
\label{outside}
Using the above decomposition, the CF of $P_{I_2}$ can be expressed as 
\begin{equation}
   \phi_{P_{I_2}}(t) = \prod\limits_{n=1}^{\infty} \;\; \exp\Bigg\{-2\pi \lambda_{R,n}\int_{r_1}^{\infty} \bigg[1 - \Big(\sum_{m=1}^{\infty} p(m) (1-jtr^{-\alpha}s_m)^{k_m} \Big)^n \bigg] rdr \Bigg\}
\end{equation}
where 
\begin{equation}
\label{truncated}
    p(m) = \dfrac{\Big[\lambda_R\pi(r_1^2-r_0^2)\Big]^{m}}{m!}\dfrac{e^{-\lambda_R\pi(r_1^2-r_0^2)}}{1 - e^{-\lambda_R\pi(r_1^2-r_0^2)}}.
\end{equation}

\proof
See appendix \ref{proof_outside}.
\end{proposition}

\section{Numerical Results}

\subsection{Interference analysis}

\begin{figure}[h!]
    \centering
    \includegraphics[width = 0.7\textwidth]{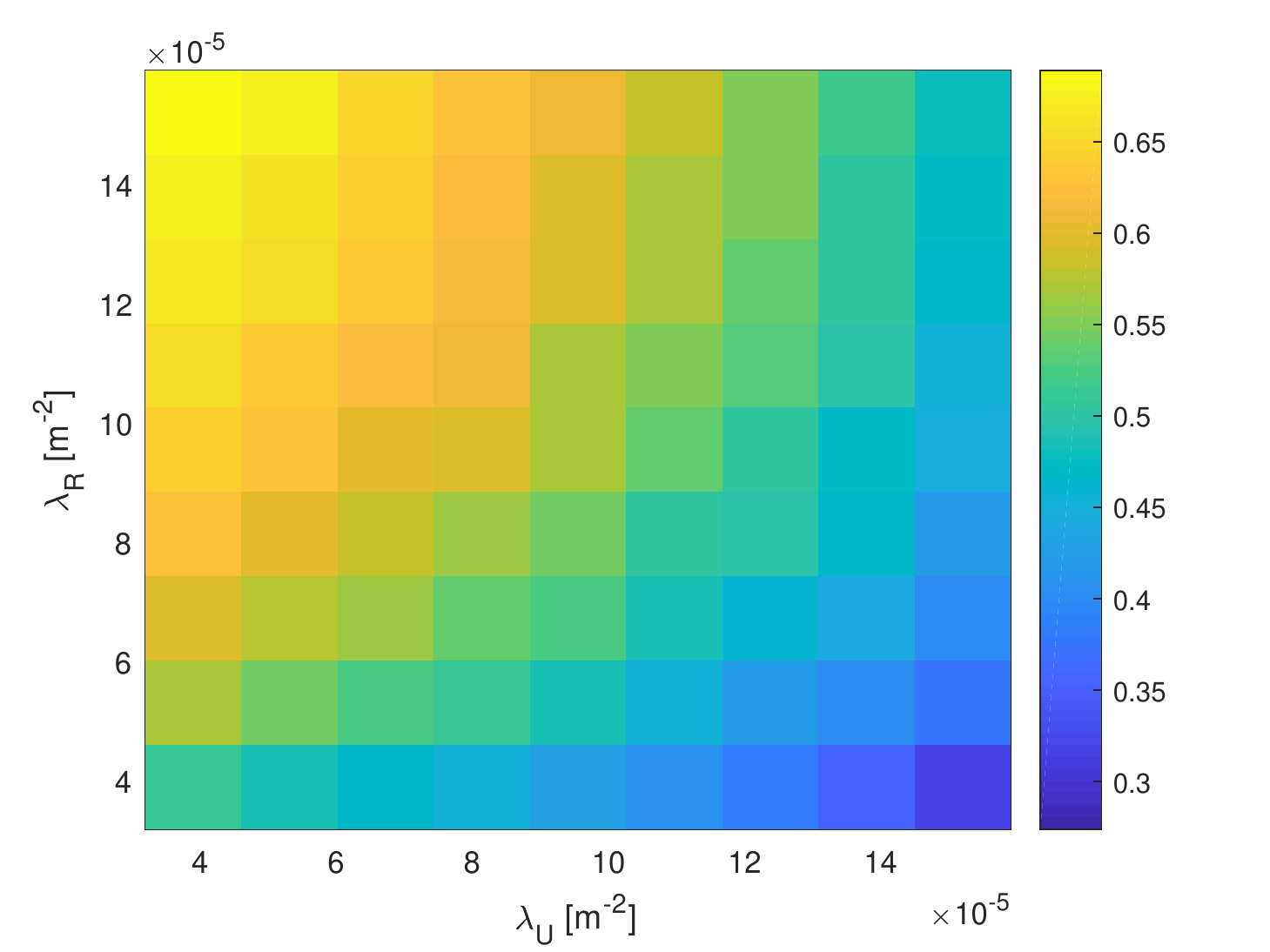}
    \caption{Ratio of the interference terms $P_{I_1}$ and $P_{I_2}$. The parameters selected were $r_1 = 100 \; m$, $N_0 = 0 \; W$, $M = 1$ and $\alpha = 3$.}
    \label{ratio_interf}
\end{figure}

Figure \ref{ratio_interf} represents the average ratio of the interference powers $P_{I_1}$ and $P_{I_2}$, under the hypotheses of Propositions 1 to 5. This ratio is represented as function of the user and RRH densities, whose values correspond here to an average number of nodes ranging from 1 to 5 per coordination zone. One can observe that the inner interference $P_{I_1}$ rapidly becomes dominant as these two parameters increases. This observation illustrates the importance of modeling this term accurately in the SG analysis.

\subsection{Validity of the stochastic geometry model and influence of the number of antennas}

\begin{figure}[h!]
    \centering
    \includegraphics[width = 0.7\textwidth]{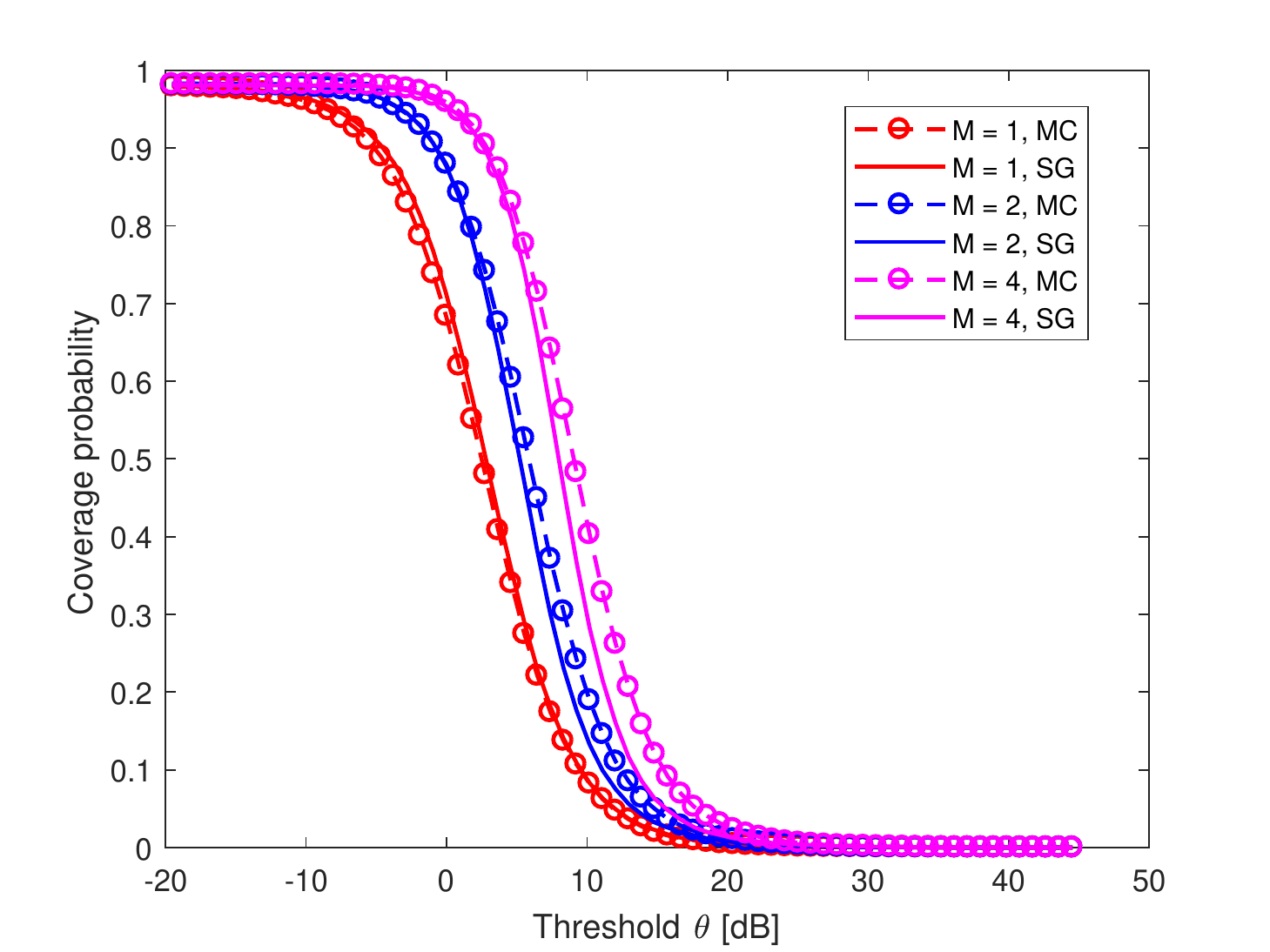}
    \caption{Impact of the number of transmit antennas on the coverage. The parameters selected here were $r_1 = 100 \; m$, $\lambda_R = 1.27 \; 10^{-4} \; m^{-2}$ (corresponding to an average of 4 RRHs per cooperative set), $\lambda_U = 3.18 \; 10^{-5} \; m^{-2}$ (corresponding to an average of 1 user per cooperative set), $N_0 = 0 \; W$ and $\alpha = 2.01$.}
    \label{antenna}
\end{figure}

An example of coverage probability curve is depicted in Figure \ref{antenna} for several antennas numbers. This coverage statistically increases with $M$ since it benefits from antenna diversity. One can also observe that the analytical values obtained by means of SG slightly differ from the values computed with the Monte Carlo simulations. This difference comes from the assumptions required in Proposition 1 to 5. 

\subsection{Spectral efficiency}

\begin{figure}
    \centering
    \includegraphics[width = 0.7\textwidth]{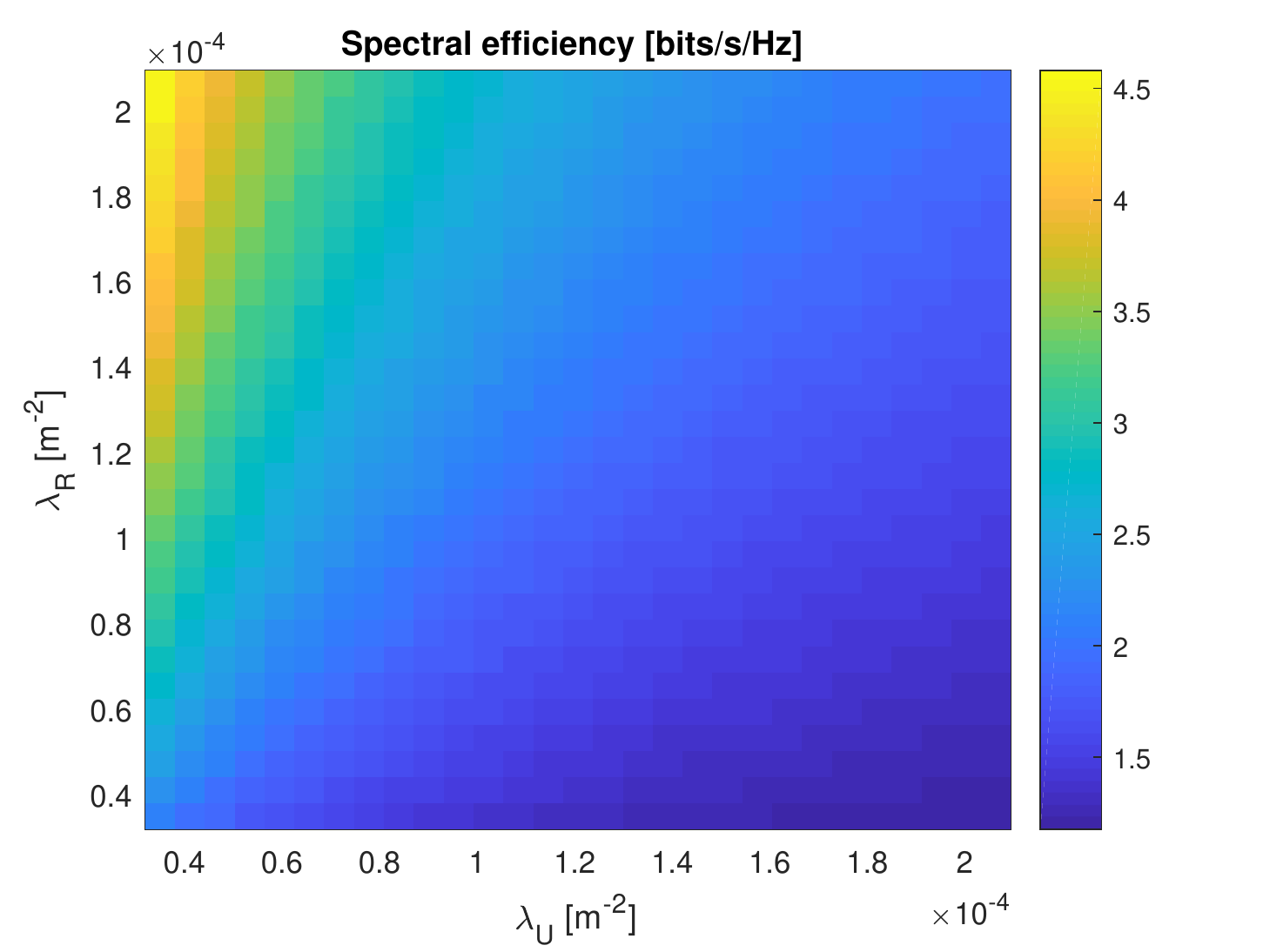}
    \caption{Joint influence of the RRH and user densities. The other parameters selected here were $M=1$, $\alpha = 3$, $N_0 = 0 \; W$ and $r_1 = 100 \; m$. The densities used in the figure therefore correspond to a range of 1 to 6.5 nodes in the cooperative sets in average. }
    \label{SE}
\end{figure}

The colormap represented in Figure \ref{SE} illustrates the variation of the average spectral efficiency as function of the user and RRH densities. As a reminder, the power allocation considered in this model is such that every user is granted a unit power distributed among the RRHs in its cooperation sets. The spectral efficiency therefore decreases with the user density owing to a higher aggregate interference. By contrast, increasing the RRH density enables to generate more nodes per cooperative set. The user data rate therefore benefits from spatial diversity thanks to the number of sources.  One can conclude that this proposed scheme is thus particularly efficient for regimes satisfying $\lambda_R >> \lambda_U$.

\section{Conclusion}
This paper considered a JT scheme developed for C-RAN. The specificities of this scheme are the real time dynamic clustering and joint power allocation. The performance metrics of the network are calculated thanks to SG. The derived expressions take into account both user and RRH densities. They also capture the correlation between the useful and interference powers, coming from the overlap of the cooperative sets. \\

Further research directions could include the incorporation of other services in the network model (e.g. wireless power transfer). Regarding the beamforming design, other normalization strategies could be studied and compared with the power allocation of this paper. The uplink channel estimation and complexity issues regarding the signal processing were also not considered in this work.

\appendices

\section{Proof of Proposition \ref{CT}}
\label{proof_CT}
By replacing the coefficients $\mathbf{w}_{ij}$ by their definition, the expression of $P_{I}$ in (\ref{eqn_power}) can be rewritten as 
\begin{equation}
            P_{I} = \sum_{j \in \Psi_{U} \backslash \{j^*\}} \frac{1}{|\mathbf{g}_{j}|^2}
            \mathbf{\bigg|} \sum_{i \in \mathcal{C}_{j}}  \mathbf{h}_{ij}^H \mathbf{h}_{ij^*} r^{-\alpha/2}_{ij}  r^{-\alpha/2}_{ij^*} \mathbf{\bigg|^2}.
\end{equation}
By developing the squared norm in this last equation, one obtains
\begin{align}
\label{proof1}
        P_{I} &= \sum_{j \in \Psi_{U} \backslash \{j^*\}} \frac{1}{|\mathbf{g}_{j}|^2}  \sum_{i \in \mathcal{C}_{j}}  \bigg| \mathbf{h}_{ij}^H \mathbf{h}_{ij^*} r^{-\alpha/2}_{ij}  r^{-\alpha/2}_{ij^*} \bigg|^2\\
        &+  \sum_{j \in \Psi_{U}} \frac{1}{|\mathbf{g}_{j}|^2} \sum_{i \in \mathcal{C}_j} \sum_{i' \in \mathcal{C}_{j}, i' \neq i} \underbrace { \mathbf{h}_{ij}^H \mathbf{h}_{ij^*} r^{-\alpha/2}_{ij}  r^{-\alpha/2}_{ij^*} \mathbf{h}_{i'j}^H \mathbf{h}_{i'j^*} r^{-\alpha/2}_{i'j}  r^{-\alpha/2}_{i'j^*} }_{\text{cross terms}}.
\end{align}

The cross terms indicated in the above equation are zero mean since the vectors $\mathbf{h}_{ij}$ contain zero mean elements $h_{ijk} \sim \mathcal{CN}\big(0,\frac{\sqrt{2}}{2}\big)$. The total three-fold sum is therefore also zero mean. Since the number of terms involved in this summation is high in our case, these cross terms are neglected, following the law of large numbers.

By rewriting the dot product $\mathbf{h}_{ij}^H \mathbf{h}_{ij^*}$, one obtains 
\begin{equation}
        P_{I} = \sum_{j \in \Psi_{U} \backslash \{j^*\}} \frac{1}{|\mathbf{g}_{j}|^2}  \sum_{i \in \mathcal{C}_{j}}  \bigg| \Big(\sum_{k=1}^{M} h_{ijk}^* h_{ij^*k}  \Big) r^{-\alpha/2}_{ij}  r^{-\alpha/2}_{ij^*} \bigg|^2.
\end{equation}

Developing the remaining squared norm and neglecting a second time the cross terms, we obtain 
\begin{equation*}
     P'_{I} = \sum_{j \in \Psi_{U} \backslash \{j^*\}} \; \sum_{i' \in \mathcal{C}_{j} }  \;  \dfrac{\sum_{k=1}^{M} \big| h_{ijk}\big|^2 \big| h_{ij^*k} |^2 r^{-\alpha}_{ij}}{\sum\limits_{i' \in \mathcal{C}_{j}} \sum_{k=1}^{M} \big| h_{i'jk}\big|^2 r^{-\alpha}_{i'j}} \; r^{-\alpha}_{ij^*}.
\end{equation*}

The two sums in the last result can then be swapped and rewritten as 
        \begin{align}
        P'_{I} &= \sum_{i \in \Psi_{R}} \; \sum_{\substack{j \in \mathcal{B}_i \\ j \neq j^*}} \;  \dfrac{\sum_{k=1}^{M} \big| h_{ijk}\big|^2 \big| h_{ij^*k} |^2 r^{-\alpha}_{ij}}{\sum_{i' \in \mathcal{C}_{j}} \sum_{k=1}^{M} \big| h_{i'jk}\big|^2 r^{-\alpha}_{i'j}} \; r^{-\alpha}_{ij^*} \\
        &= \sum_{i \in \mathcal{C}_{j^*}} \; \sum_{\substack{j \in \mathcal{B}_i \\ j \neq j^*}} \;  \dfrac{\sum_{k=1}^{M} \big| h_{ijk}\big|^2 \big| h_{ij^*k} |^2 r^{-\alpha}_{ij}}{\sum_{i' \in \mathcal{C}_{j}} \sum_{k=1}^{M} \big| h_{i'jk}\big|^2 r^{-\alpha}_{i'j}} \; r^{-\alpha}_{ij^*} + \sum_{i \in \Psi_{R}\setminus \mathcal{C}_{j^*}} \; \sum_{\substack{j \in \mathcal{B}_i \\ j \neq j^*}} \;  \dfrac{\sum_{k=1}^{M} \big| h_{ijk}\big|^2 \big| h_{ij^*k} |^2 r^{-\alpha}_{ij}}{\sum_{i' \in \mathcal{C}_{j}} \sum_{k=1}^{M} \big| h_{i'jk}\big|^2 r^{-\alpha}_{i'j}} \; r^{-\alpha}_{ij^*}
        \end{align}

where $\mathcal{B}_i$ denotes the set of users served by RRH $i$.

\section{Proof of Proposition \ref{moment_matching}}
\label{proof_moment_matching}
\begin{figure}[h!]
    \centering
    \includegraphics[width = 0.7\textwidth]{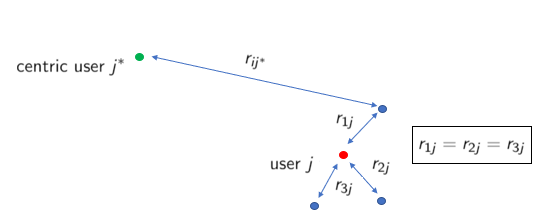}
    \caption{Illustration of the distance approximation: the three RRHs serving user $j$ are assumed to be located at the same distance from it.}
    \label{dist_approx}
\end{figure}

The initial expression is given by 
\begin{equation}
\label{fraction_moment}
\dfrac{\sum_{k=1}^{M} \big| h_{ijk}\big|^2 \big| h_{ij^*k} |^2 r^{-\alpha}_{ij}}{\sum_{i'=1}^{N_j} \sum_{k=1}^{M} \big| h_{i'jk}\big|^2 r^{-\alpha}_{i'j}} 
\end{equation}

in this quotient, the distances $r_{ij}$ and $r_{i'j}$ represent the distances from two base stations serving a user $j$ (different from the centric user). In order to derive a tractable expression for the distribution of (\ref{fraction_moment}), it is assumed that the distances from the RRH serving a user $j \neq j^*$ are equal. This assumption is illustrated in Figure \ref{dist_approx}.

The expression left is 
\begin{equation}
\label{fraction_moment2}
Z \triangleq \dfrac{X}{Y} = \dfrac{\sum_{k=1}^{M} \big| h_{ijk}\big|^2 \big| h_{ij^*k} |^2} {\sum_{i'=1}^{N_j} \sum_{k=1}^{M} \big| h_{i'jk}\big|^2}. 
\end{equation}

The distribution of $Z$ is approximated by a Gamma distribution using moment matching. In order to compute the shape and scale parameters of the Gamma distribution, it is necessary to compute the mean and the variance of $Z$. Using the Taylor approximations presented in \cite{ratio_1,ratio_2,ratio_3}, it is possible to compute these moments based on the moments of the numerator $X$ and $Y$.The shape and scale parameters are then given by $k = E[Z]^2/\text{Var}[Z]$ and $s = \text{Var}[Z]/E[Z]$.

\section{Proof of Proposition \ref{GP_corr}}
\label{proof_GP_corr}
Let $n_U$ and $n_R$ be the number of users and RRH in the cooperative set of $j^*$ (i.e. located at a distance in $[r_0;r_1]$). Conditioned on these numbers, the coverage probability can be expressed as
\begin{equation}
\begin{aligned}
    \mathcal{P}(\theta|n_U,n_R) &\triangleq \mathbb{P}\Big[\text{SINR}>\theta |n_U,n_R \Big] \\
    &\stackrel{(a)}{=} \mathbb{P}\Big[P_U - \theta P_I - \theta N_0 > 0|n_U,n_R \Big]\\
    &\stackrel{(b)}{=} \dfrac{1}{2} +\dfrac{1}{\pi}\int_{0}^{\infty}\dfrac{1}{t}\text{Im}\bigg[\phi_{P_U}(t|n_U,n_R)\phi_{P_I}(-t\theta|n_U,n_R)e^{jt\theta N_0}\bigg]dt \\
    &\stackrel{(c)}{=} \dfrac{1}{2} +\dfrac{1}{\pi}\int_{0}^{\infty}\dfrac{1}{t}\text{Im}\bigg[\underbrace{\phi_{P_U}(t|n_U,n_R)\phi_{P_{I_1}}(-t\theta|n_U,n_R)}_{}\phi_{P_{I_2}}(-t\theta|n_U,n_R)e^{jt\theta N_0}\bigg]dt \\
    &\stackrel{(d)}{=} \dfrac{1}{2} +\dfrac{1}{\pi}\int_{0}^{\infty}\dfrac{1}{t}\text{Im}\bigg[\phi_{P_{U'}}(t|n_U,n_R)\phi_{P_{I_2}}(-t\theta)e^{jt\theta N_0}\bigg]dt \\
\end{aligned}
\end{equation}
where (a) comes from the definition of the SINR, (b) is obtained by applying Gil-Pelaez theorem \cite{GP}, (c) comes from the decomposition of the total interference into $I_1$ and $I_2$ and (d) is obtained by defining $P_{U'} = P_{U} - \theta P_{I_2}$ and grouping the two related CFs. Note that $\phi_{P_{I_2}}$ is assumed to be independent from $n_R$ and $n_U$\footnote{to be rigorous, there actually exist correlation between $P_{I_2}$ and $P_{U}$: the RRHs located at a distance slightly larger than $r_1$ will serve users located inside $[r_0;r_1]$, and counted in $n_U$. Compared to $P_{I_1}$, this correlation is however much lower, and is therefore neglected here.}.

The total coverage is obtained by deconditioning over $n_U$ and $n_R$ 
\begin{equation}
    \mathcal{P}(\theta) = \sum_{n_U = 0}^{\infty}\sum_{n_R = 0}^{\infty} p(n_R)p(n_U) \mathcal{P}(\theta|n_U,n_R)
\end{equation}
As a reminder, $\Psi_R$ and $\Psi_U$ are both HPPP. $p(n_R)$ and $p(n_U)$ are therefore the probability mass functions of Poisson random variables, with of respective parameters $\lambda_R\pi(r_1^2-r_0^2)$ and $\lambda_U\pi(r_1^2-r_0^2)$.

\section{Proof of Proposition \ref{corr2}}
\label{proof_corr2}
The variable $P_{U'}$ can be decomposed in the following manner
\begin{equation*}
\begin{aligned}
    P_{U'} = P_U - \theta P_{I_1} = \sum_{i=1}^{n_R} P_{U,i} - \theta \sum_{i=1}^{n_R} P_{I_1,i} = \sum_{i=1}^{n_R}  \Big(P_{U,i}- \theta P_{I_1,i} \big)
\end{aligned}
\end{equation*}
where $P_{U,i}$ and $P_{I_1,i}$ are the useful and interference power coming from RRH $i$. The functions $\phi_S(\cdot)$ and $\phi_T(\cdot)$ in (\ref{SandT}) are the CFs of $P_{U,i}$ and $P_{I_1,i}$, conditioned on the distance between $i$ and the centric user. The characteristic function of $P_{U,i}- \theta P_{I_1,i}$ is therefore given by $\phi_{S}(t|n_R,n_U,r)\phi_{T}(-\theta t|n_R,n_U,r)$. This CF is then integrated over the pdf of the distance $r$ from the origin, given by
\begin{equation}
    f_r(r) = \frac{2r}{r_1^2-r_0^2} \; \; \; \text{with} \; \; \; r_0 < r < r_1
\end{equation}

This leads to equation \ref{SandT}. The rest of this proof consists in deriving $\phi_S(\cdot)$ and $\phi_T(\cdot)$.

By replacing $\mathbf{w}_{ij}$ by its definition in (\ref{eqn_power}), we obtain 
\begin{equation}
    P_U = \sum_{i \in \mathcal{C}_j} |\mathbf{h}_{ij^*}|^2r_{ij^*}^{-\alpha}.
\end{equation}

$\phi_S(\cdot)$ is the CF of one term of this last expression. Each of these terms follows a Gamma distribution of shape $M$ and scale $r_{ij^*}^{-\alpha}$, which results in (\ref{phi_S}).

The second expression to compute is $\phi_T(\cdot)$. As a reminder, the interference $P_{I_1}$ was expressed in (\ref{eqn_power}) as 
\begin{equation}
   P_{I_1} = \sum_{i \in \mathcal{C}_{j^*}} \; \underbrace{\sum_{\substack{j \in \mathcal{B}_i \\ j \neq j^*}} \;  Z^{(i,j)}_{|\mathcal{C}_{j}|} r^{-\alpha}_{ij^*}}_{W_i}.
\end{equation}

$\phi_T(\cdot)$ is the CF of one term $W_i$ coming from a RRH $i$. 

Let $n_U' = |\mathcal{B}_i|$ and $n_R' = |\mathcal{C}_j|$ be the number of RRH serving a user $j \in \mathcal{B}_i$. Conditioned on this latter number and on the distance, the variable $Z^{(i,j)}_{n_R'} r^{-\alpha}_{ij^*}$ follows a Gamma distribution. The shape and scale parameters of this distribution are given by $k_{n'_R}$ and $s_{n'_R}r^{-\alpha}_{ij^*}$. The characteristic function of this distribution is given by (\ref{phiV}).

By averaging over $n_U'$ and $n_R'$, the expression of $\phi_T(\cdot)$ in (\ref{phi_T}) is obtained. 

\begin{figure}[h!]
    \centering
    \includegraphics[width = 0.7 \textwidth]{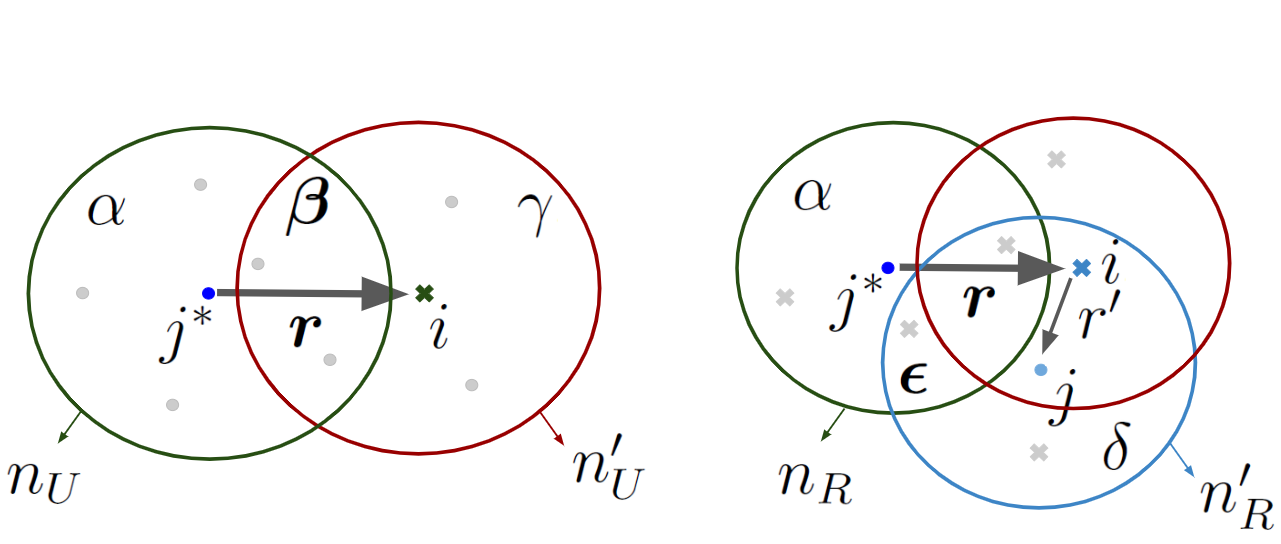}
    \caption{Intersections considered to compute probabilities $p(n_U'|n_U)$ and $p(n_R'|n_R)$.}
    \label{proof_circles_corr}
\end{figure}

The probabilities $p(n_U'|n_U)$ and $p(n_R'|n_R)$ can be computed with the help of Figure \ref{proof_circles_corr}.

In this figure, the area of the intersection $\beta$ of the two disks around $j^*$ and $i^*$ is given by \cite{circle_intersect}:
\begin{equation}
    \mathcal{A}(\beta) = 2r_1^{2} \text{acos}\Big(\dfrac{r}{2r_1} \Big) - \dfrac{r}{2}\sqrt{4r_1^2-r^2}.
\end{equation}

In order to obtain a simpler form, the next results are derived using the linearization around zero of this expression: 
\begin{equation}
    \Tilde{\mathcal{A}}(\beta) = \pi r_1^2 \bigg[1 - \frac{2r}{\pi r_1} \bigg].
\end{equation}

Since the users are distributed using a PPP, the number of nodes in the zones $\alpha$, $\beta$ and $\gamma$ follow independent Poisson distributions. 

The conditional probability of $n_U'$ is hence given by
\begin{align}
    p(n_U'|n_U) &= \dfrac{p\bigg[\Big(n_U \; \text{in} \; (\alpha \cup \beta) \Big) \cap \Big(n_U' \; \text{in} \; (\beta \cup \gamma) \Big) \bigg]}{p\Big[n_U \; \text{in} \;  (\alpha \cup \beta) \Big]} \\
    &= \dfrac{\sum_{k=0}^{\min(n_U,n_U')} p\big[n_U - k \; \text{in} \;  \alpha \big]p\big[k \; \text{in} \;  \beta \big]p\big[n_U'-k \; \text{in} \;  \gamma \big]}{p\big[n_U \; \text{in} \;  (\alpha \cup \beta) \big]}
\end{align}
where 
\begin{align*}
    p\big[n_U - k \; \text{in} \;  \alpha \big] &= \bigg[\lambda_U \pi r_1^2\Big(1 - \frac{2r}{\pi r_1} \Big) \bigg]^{n_U-k} \exp\bigg[- \lambda_U \pi r_1^2\Big(1 - \frac{2r}{\pi r_1} \Big)\bigg] \bigg[(n_U-k)!\bigg]^{-1}\\
    p\big[k \; \text{in} \;  \beta \big] &= \big(\lambda_U\pi r_1^2 \big)^k \exp\big(-\lambda_U\pi r_1^2  \big) \big(k! \big)^{-1}\\
    p\big[n_U'-k \; \text{in} \;  \gamma \big] &= \bigg[\lambda_U \pi r_1^2\Big(1 - \frac{2r}{\pi r_1} \Big) \bigg]^{n_U'-k} \exp\bigg[- \lambda_U \pi r_1^2\Big(1 - \frac{2r}{\pi r_1} \Big)\bigg] \bigg[(n_U'-k)!\bigg]^{-1}\\
    p\big[n_U \; \text{in} \;  (\alpha \cup \beta) \big] &= \big(\lambda_U \pi r_1^2 \big)^{n_U} \exp\big(- \lambda_U \pi r_1^2\big) \big(n_U! \big)^{-1}.
\end{align*}

The probability of $n_R$ is hence given by
\begin{align}
    \label{pp}
    p(n_R'|n_R) &= \dfrac{p\bigg[\Big(n_R \; \text{in} \; (\alpha \cup \epsilon) \Big) \cap \Big(n_R' \; \text{in} \; (\epsilon \cup \delta) \Big) \Big| n_R>1 \bigg]}{p\Big[n_R' \; \text{in} \;  (\alpha \cup \epsilon) \Big| n_R>1 \Big]} \\
    &= \dfrac{\sum_{k=1}^{\min(n_R,n_R')} p\big[n_R - k \; \text{in} \;  \alpha \big]p\big[k \; \text{in} \;  \epsilon \big]p\big[n_R'-k \; \text{in} \;  \delta \big]}{\sum_{k=1}^{n_R} p\big[k \; \text{in} \;  \epsilon \big]p\big[n_R-k \; \text{in} \;  \alpha \big]}.
\end{align}

\noindent
The probabilities involved in this last expression require to compute the area of the intersection $\epsilon$ in Figure \ref{proof_circles_corr}. This area depends on the distance between the green and blue circle, given by $r'' = \big(r^2+r'^2-2rr'\cos(\angle r r')\big)^{1/2}$. The resulting intersection should be average over the distribution of $r'$, which does not lead to tractable expressions. In order to work with a simple expression, we consider an area of intersection given by a linear function $\chi + \zeta r$, whose values are determined by considering the average case $r'=r_1/2$ and $\angle r r' = \pi /2$. The resulting values are given by 
\begin{align}
    \chi &= \bigg(2\text{acos}\Big(\frac{\sqrt{5}}{4}\Big) - \frac{5}{4} \sqrt{4-\frac{1}{4}}\bigg) \\
    \zeta &= \frac{1}{r_1}\bigg(\chi - 2\text{acos}\Big(\frac{1}{4}\Big) + \frac{1}{4} \sqrt{4-\frac{1}{4}}\bigg).
\end{align}

The probabilities present in \ref{pp} can hence be expressed as 
\begin{align*}
    p\big[n_R - k \; \text{in} \;  \alpha \big] &= \bigg[\lambda_R \pi r_1^2\Big(\chi + \zeta r \Big) \bigg]^{n_R-k} \exp\bigg[- \lambda_U \pi r_1^2\Big(\chi + \zeta r \Big)\bigg] \bigg[(n_R-k)!\bigg]^{-1}\\
    p\big[k \; \text{in} \;  \epsilon \big] &= \big(\lambda_R\pi r_1^2 \big)^k \exp\big(-\lambda_R\pi r_1^2  \big) \big(k! \big)^{-1}\\
    p\big[n_R'-k \; \text{in} \;  \delta \big] &= \bigg[\lambda_R \pi r_1^2\Big(\chi + \zeta r\Big) \bigg]^{n_R'-k} \exp\bigg[- \lambda_R \pi r_1^2\Big(\chi + \zeta r \Big)\bigg] \bigg[(n_R'-k)!\bigg]^{-1}\\
    p\big[n_R \; \text{in} \;  (\epsilon \cup \delta) \big] &= \big(\lambda_U \pi r_1^2 \big)^{n_R} \exp\big(- \lambda_R \pi r_1^2\big) \big(n_R! \big)^{-1}\bigg[\chi + \zeta r\bigg].
\end{align*}

\section{Proof of Proposition \ref{outside}}
\label{proof_outside}
On the basis of the proposed decomposition, the interference $P_{I_2}$ can be expressed as 
\begin{equation}
    P_{I_2} = \sum_{n=1}^{\infty} \; \underbrace{\sum_{i \in \Tilde{\Psi}_{R,n}} \; \sum_{j=1}^{n} \;  Z^{(i,j)}_{|\mathcal{C}_{j}|} r^{-\alpha}_{ij^*}}_{P_{I_2,n}}.
\end{equation}

where $P_{I_2,n}$ is the interference coming from the PPP $\Tilde{\Psi}_{R,n}$.
The CF of $P_{I_2}$ can hence be expressed as 
\begin{equation}
    \phi_{P_{I_2}}(t) = \prod_{n=1}^{\infty} \phi_{P_{I_2,n}}(t).
\end{equation}
with $\phi_{P_{I_2,n}}(t)$, the CF of the interference coming from $\Tilde{\Psi}_{R,n}$. 

Let us introduce the slack variable $Y_{i,n} = \sum_{j=1}^{n} \;  Z^{(i,j)}_{|\mathcal{C}_{j}|}$. The CF can be calculated as follows 

\begin{equation}
\begin{aligned}
\phi_{P_{I_2,n}}(t) &= \mathbb{E}_{\Tilde{\Psi}_{R,n}} \Bigg\{ \prod_{i \in \Tilde{\Psi}_{R,n}} \mathbb{E}_{Y_{i,n}}\Big[ \exp \big(jtY_{i,n}r^{-\alpha}_{ij^*} \big) \Big] \Bigg\} \\
&\stackrel{(a)}{=} \mathbb{E}_{\Tilde{\Psi}_{R,n}} \Bigg\{ \prod_{i \in \Tilde{\Psi}_{R,n}} \phi_{Y_{i,n}}(r^{-\alpha}_{ij^*} t) \Bigg\} \\
&\stackrel{(b)}{=} \exp\Bigg\{-2\pi \lambda_{R,n}\int_{r_1}^{\infty} \Big[1 - \phi_{Y_{i,n}}(r^{-\alpha} t) \Big] rdr \Bigg\}
\end{aligned}
\end{equation}

where (a) comes from the definition of the CF and (b) is obtained using the probability generating functional (PGFL) of a PPP \cite{Baccelli}. 

$\phi_{Y_{i,n}}(t)$ is the CF of $\sum_{j=1}^{n} \;  Z^{(i,j)}_{|\mathcal{C}_{j}|}$. All the $n$ terms of this summation have the same distribution. This distribution has to be averaged over the possible values of $|\mathcal{C}_{j}|$. As a result, we have
\begin{equation}
    \phi_{Y_{i,n}}(t) = \Big(\sum_{m=1}^{\infty} p(m) (1-jts_m)^{k_m} \Big)^n
\end{equation}

where $p(m)$ is given by (\ref{truncated}). Note that we here use a truncated Poisson distribution since the each considered user is served at least by one RRH.

\section*{Acknowledgment}

This work was supported by F.R.S.-FNRS under the EOS program (EOS project 30452698).

\ifCLASSOPTIONcaptionsoff
  \newpage
\fi

\bibliographystyle{IEEEtran}
\bibliography{References}

\begin{thebibliography}{10}
\providecommand{\url}[1]{#1}
\csname url@samestyle\endcsname
\providecommand{\newblock}{\relax}
\providecommand{\bibinfo}[2]{#2}
\providecommand{\BIBentrySTDinterwordspacing}{\spaceskip=0pt\relax}
\providecommand{\BIBentryALTinterwordstretchfactor}{4}
\providecommand{\BIBentryALTinterwordspacing}{\spaceskip=\fontdimen2\font plus
\BIBentryALTinterwordstretchfactor\fontdimen3\font minus
  \fontdimen4\font\relax}
\providecommand{\BIBforeignlanguage}[2]{{%
\expandafter\ifx\csname l@#1\endcsname\relax
\typeout{** WARNING: IEEEtran.bst: No hyphenation pattern has been}%
\typeout{** loaded for the language `#1'. Using the pattern for}%
\typeout{** the default language instead.}%
\else
\language=\csname l@#1\endcsname
\fi
#2}}
\providecommand{\BIBdecl}{\relax}
\BIBdecl

\bibitem{Intro_1}
M.~{Agiwal}, A.~{Roy}, and N.~{Saxena}, ``Next generation 5g wireless networks:
  A comprehensive survey,'' \emph{IEEE Communications Surveys Tutorials},
  vol.~18, no.~3, pp. 1617--1655, 2016.

\bibitem{9186615}
K.~{Ntontin} and C.~{Verikoukis}, ``System-level analysis of a
  self-fronthauling and millimeter-wave cloud-ran,'' \emph{IEEE Transactions on
  Communications}, vol.~68, no.~12, pp. 7762--7778, 2020.

\bibitem{7491366}
J.~{Park} and R.~W. {Heath}, ``Low complexity antenna selection for low target
  rate users in dense cloud radio access networks,'' \emph{IEEE Transactions on
  Wireless Communications}, vol.~15, no.~9, pp. 6022--6032, 2016.

\bibitem{7944682}
C.~{Skouroumounis}, C.~{Psomas}, and I.~{Krikidis}, ``Low-complexity base
  station selection scheme in mmwave cellular networks,'' \emph{IEEE
  Transactions on Communications}, vol.~65, no.~9, pp. 4049--4064, 2017.

\bibitem{6600939}
Z.~{Ding} and H.~V. {Poor}, ``The use of spatially random base stations in
  cloud radio access networks,'' \emph{IEEE Signal Processing Letters},
  vol.~20, no.~11, pp. 1138--1141, 2013.

\bibitem{7018059}
Z.~{Yang}, Z.~{Ding}, and P.~{Fan}, ``Performance analysis of cloud radio
  access networks with uniformly distributed base stations,'' \emph{IEEE
  Transactions on Vehicular Technology}, vol.~65, no.~1, pp. 472--477, 2016.

\bibitem{7124520}
F.~A. {Khan}, H.~{He}, J.~{Xue}, and T.~{Ratnarajah}, ``Performance analysis of
  cloud radio access networks with distributed multiple antenna remote radio
  heads,'' \emph{IEEE Transactions on Signal Processing}, vol.~63, no.~18, pp.
  4784--4799, 2015.

\bibitem{7415954}
H.~{He}, J.~{Xue}, T.~{Ratnarajah}, F.~A. {Khan}, and C.~B. {Papadias},
  ``Modeling and analysis of cloud radio access networks using matérn
  hard-core point processes,'' \emph{IEEE Transactions on Wireless
  Communications}, vol.~15, no.~6, pp. 4074--4087, 2016.

\bibitem{8400544}
M.~{Mohammadi}, H.~A. {Suraweera}, and C.~{Tellambura}, ``Uplink/downlink rate
  analysis and impact of power allocation for full-duplex cloud-rans,''
  \emph{IEEE Transactions on Wireless Communications}, vol.~17, no.~9, pp.
  5774--5788, 2018.

\bibitem{9217353}
A.~{Papazafeiropoulos}, H.~Q. {Ngo}, P.~{Kourtessis}, S.~{Chatzinotas}, and
  J.~M. {Senior}, ``Optimal energy efficiency in cell-free massive mimo
  systems: A stochastic geometry approach,'' in \emph{2020 IEEE 31st Annual
  International Symposium on Personal, Indoor and Mobile Radio Communications},
  2020, pp. 1--7.

\bibitem{8972478}
A.~{Papazafeiropoulos}, P.~{Kourtessis}, M.~D. {Renzo}, S.~{Chatzinotas}, and
  J.~M. {Senior}, ``Performance analysis of cell-free massive mimo systems: A
  stochastic geometry approach,'' \emph{IEEE Transactions on Vehicular
  Technology}, vol.~69, no.~4, pp. 3523--3537, 2020.

\bibitem{7334848}
F.~{Ghods}, A.~O. {Fapojuwo}, and F.~{Ghannouchi}, ``Energy efficiency and
  spectrum efficiency in cooperative cloud radio access network,'' in
  \emph{2015 IEEE Pacific Rim Conference on Communications, Computers and
  Signal Processing (PACRIM)}, 2015, pp. 280--285.

\bibitem{7425275}
J.~{Liu}, M.~{Sheng}, T.~Q.~S. {Quek}, and J.~{Li}, ``D2d enhanced co-ordinated
  multipoint in cloud radio access networks,'' \emph{IEEE Transactions on
  Wireless Communications}, vol.~15, no.~6, pp. 4248--4262, 2016.

\bibitem{7460258}
S.~{Zhan} and D.~{Niyato}, ``A coalition formation game for remote radio head
  cooperation in cloud radio access network,'' \emph{IEEE Transactions on
  Vehicular Technology}, vol.~66, no.~2, pp. 1723--1738, 2017.

\bibitem{7982774}
M.~{Cheng}, J.~{Wang}, Y.~{Wu}, and M.~{Lin}, ``Downlink ergodic rate analysis
  for virtual cell based cloud radio access networks,'' \emph{IEEE Access},
  vol.~5, pp. 13\,520--13\,530, 2017.

\bibitem{8063963}
X.~{Gu}, X.~{Ji}, Z.~{Ding}, W.~{Wu}, and M.~{Peng}, ``Outage probability
  analysis of non-orthogonal multiple access in cloud radio access networks,''
  \emph{IEEE Communications Letters}, vol.~22, no.~1, pp. 149--152, 2018.

\bibitem{8422228}
U.~S. {Hashmi}, S.~A.~R. {Zaidi}, A.~{Darbandi}, and A.~{Imran}, ``On the
  efficiency tradeoffs in user-centric cloud ran,'' in \emph{2018 IEEE
  International Conference on Communications (ICC)}, 2018, pp. 1--7.

\bibitem{9024639}
Q.~{Zhu}, X.~{Wang}, and Z.~{Qian}, ``An analytical framework for clustering
  mechanism with nakagami fading in user-centric cloud ran,'' in \emph{2019
  IEEE Globecom Workshops (GC Wkshps)}, 2019, pp. 1--6.

\bibitem{9082905}
S.~M. {Azimi-Abarghouyi}, M.~{Nasiri-Kenari}, and M.~{Debbah}, ``Stochastic
  design and analysis of user-centric wireless cloud caching networks,''
  \emph{IEEE Transactions on Wireless Communications}, vol.~19, no.~7, pp.
  4978--4993, 2020.

\bibitem{8254187}
F.~J. {Martin-Vega}, Y.~{Liu}, G.~{Gomez}, M.~C. {Aguayo-Torres}, and
  M.~{Elkashlan}, ``Modeling and analysis of noma enabled cran with cluster
  point process,'' in \emph{GLOBECOM 2017 - 2017 IEEE Global Communications
  Conference}, 2017, pp. 1--6.

\bibitem{9050646}
M.~{Elhattab} and W.~{Hamouda}, ``Performance analysis for h-crans under
  constrained capacity fronthaul,'' \emph{IEEE Networking Letters}, vol.~2,
  no.~2, pp. 62--66, 2020.

\bibitem{7248933}
Z.~{Zhao}, M.~{Peng}, Z.~{Ding}, C.~{Wang}, and H.~V. {Poor}, ``Cluster
  formation in cloud-radio access networks: Performance analysis and algorithms
  design,'' in \emph{2015 IEEE International Conference on Communications
  (ICC)}, 2015, pp. 3903--3908.

\bibitem{8974591}
S.~{Mukherjee} and J.~{Lee}, ``Edge computing-enabled cell-free massive mimo
  systems,'' \emph{IEEE Transactions on Wireless Communications}, vol.~19,
  no.~4, pp. 2884--2899, 2020.

\bibitem{7365885}
M.~A. {Abana}, S.~{Yaohua}, M.~{Ahmed}, L.~A. {Olawoyin}, and L.~{Yong},
  ``Performance analysis in cloud radio access networks: user-centralized
  coordination approach,'' \emph{China Communications}, vol.~12, no.~11, pp.
  1--12, 2015.

\bibitem{6860252}
R.~{Tanbourgi}, S.~{Singh}, J.~G. {Andrews}, and F.~K. {Jondral}, ``A tractable
  model for noncoherent joint-transmission base station cooperation,''
  \emph{IEEE Transactions on Wireless Communications}, vol.~13, no.~9, pp.
  4959--4973, 2014.

\bibitem{9201540}
S.~{Kusaladharma}, W.~P. {Zhu}, W.~{Ajib}, and G.~A.~A. {Baduge}, ``Stochastic
  geometry based performance characterization of swipt in cell-free massive
  mimo,'' \emph{IEEE Transactions on Vehicular Technology}, vol.~69, no.~11,
  pp. 13\,357--13\,370, 2020.

\bibitem{6928420}
G.~{Nigam}, P.~{Minero}, and M.~{Haenggi}, ``Coordinated multipoint joint
  transmission in heterogeneous networks,'' \emph{IEEE Transactions on
  Communications}, vol.~62, no.~11, pp. 4134--4146, 2014.

\bibitem{7738565}
S.~T. {Veetil}, K.~{Kuchi}, and R.~K. {Ganti}, ``Coverage analysis of cloud
  radio networks with finite clustering,'' \emph{IEEE Transactions on Wireless
  Communications}, vol.~16, no.~1, pp. 594--606, 2017.

\bibitem{8100895}
X.~{Yu}, Q.~{Cui}, and M.~{Haenggi}, ``Coherent joint transmission in downlink
  heterogeneous cellular networks,'' \emph{IEEE Wireless Communications
  Letters}, vol.~7, no.~2, pp. 274--277, 2018.

\bibitem{8937506}
S.~{Chen}, X.~{Liu}, T.~{Zhao}, H.~{Chen}, and W.~{Meng}, ``Performance
  analysis of joint transmission schemes in ultra-dense networks – a unified
  approach,'' \emph{IEEE/ACM Transactions on Networking}, vol.~28, no.~1, pp.
  154--167, 2020.

\bibitem{8690607}
J.~{Chen}, K.~S. {Liu}, and S.~{Su}, ``Performance of network-centric
  clustering for coordinated joint transmission with irregular cluster
  topology,'' in \emph{2018 IEEE 88th Vehicular Technology Conference
  (VTC-Fall)}, 2018, pp. 1--5.

\bibitem{6582745}
P.~{Xia}, C.~{Liu}, and J.~G. {Andrews}, ``Downlink coordinated multi-point
  with overhead modeling in heterogeneous cellular networks,'' \emph{IEEE
  Transactions on Wireless Communications}, vol.~12, no.~8, pp. 4025--4037,
  2013.

\bibitem{6824977}
G.~{Nigam}, P.~{Minero}, and M.~{Haenggi}, ``Coordinated multipoint in
  heterogeneous networks: A stochastic geometry approach,'' in \emph{2013 IEEE
  Globecom Workshops (GC Wkshps)}, 2013, pp. 145--150.

\bibitem{6851166}
N.~{Lee}, D.~{Morales-Jimenez}, A.~{Lozano}, and R.~W. {Heath}, ``Spectral
  efficiency of dynamic coordinated beamforming: A stochastic geometry
  approach,'' \emph{IEEE Transactions on Wireless Communications}, vol.~14,
  no.~1, pp. 230--241, 2015.

\bibitem{7414099}
J.~{Liu}, M.~{Sheng}, T.~Q.~S. {Quek}, and J.~{Li}, ``Comp transmission in
  cloud radio access networks,'' in \emph{2015 IEEE Globecom Workshops (GC
  Wkshps)}, 2015, pp. 1--6.

\bibitem{7450690}
K.~{Hosseini}, W.~{Yu}, and R.~S. {Adve}, ``A stochastic analysis of network
  mimo systems,'' \emph{IEEE Transactions on Signal Processing}, vol.~64,
  no.~16, pp. 4113--4126, 2016.

\bibitem{8289331}
M.~{Al-Saedy}, H.~{Al-Raweshidy}, H.~{Al-Hmood}, and F.~{Haider}, ``Coverage
  and effective capacity in downlink mimo multicell networks with power
  control: Stochastic geometry modelling,'' \emph{IEEE Access}, vol.~6, pp.
  9173--9185, 2018.

\bibitem{8422715}
W.~{Sun} and J.~{Liu}, ``A stochastic geometry analysis of comp-based uplink in
  ultra-dense cellular networks,'' in \emph{2018 IEEE International Conference
  on Communications (ICC)}, 2018, pp. 1--6.

\bibitem{8526351}
M.~{Farahmand} and A.~{Mohammadi}, ``Energy-efficient sparse beamforming in
  cloud radio access networks,'' \emph{Canadian Journal of Electrical and
  Computer Engineering}, vol.~41, no.~3, pp. 151--159, 2018.

\bibitem{9247173}
S.~{Fang}, G.~{Chen}, X.~{Xu}, S.~{Han}, and J.~{Tang}, ``Millimeter-wave
  coordinated beamforming enabled cooperative network: A stochastic geometry
  approach,'' \emph{IEEE Transactions on Communications}, pp. 1--1, 2020.

\bibitem{7448831}
W.~{Nie}, F.~{Zheng}, X.~{Wang}, W.~{Zhang}, and S.~{Jin}, ``User-centric
  cross-tier base station clustering and cooperation in heterogeneous networks:
  Rate improvement and energy saving,'' \emph{IEEE Journal on Selected Areas in
  Communications}, vol.~34, no.~5, pp. 1192--1206, 2016.

\bibitem{9145416}
A.~M. {Kundu} and T.~V. {Sreejith}, ``Full duplex cloud radio access networks:
  Performance gains,'' in \emph{2020 IEEE International Conference on
  Communications Workshops (ICC Workshops)}, 2020, pp. 1--6.

\bibitem{Baccelli}
B.~B. François~Baccelli, \emph{Stochastic Geometry and Wireless Networks,
  Volume I -Theory}.\hskip 1em plus 0.5em minus 0.4em\relax Foundations and
  Trends in Networking, 2009, vol.~3, no. 3-4.

\bibitem{ratio_1}
H.~Seltman, ``Approximations for mean and variance of a ratio,'' available at
  stat.cmu.edu.

\bibitem{ratio_2}
K.~O. Alan~Stuart, \emph{Kendall’s Advanced Theory of Statistics},
  6th~ed.\hskip 1em plus 0.5em minus 0.4em\relax Arnold London, 1998, vol.~1.

\bibitem{ratio_3}
N.~L.~J. Regina C. Elandt-Johnson, \emph{Survival Models and Data
  Analysis}.\hskip 1em plus 0.5em minus 0.4em\relax John Wiley and Sons NY,
  1980.

\bibitem{GP}
J.~Gil-Pelaez, ``Note on the inversion theorem,'' \emph{Biometrika}, vol.~38,
  pp. 481--482, 1951.

\bibitem{circle_intersect}
E.~Weisstein, ``Circle-circle intersection,'' available at
  mathworld.wolfram.com/Circle-CircleIntersection.html.

\end{thebibliography}

\end{document}